# Characterization of a novel plastic scintillation detector for in vivo electron dosimetry

Cornelius J. Bauer[1,*], Frank Schneider[1], Ida D. Göbel[1], Hans Oppitz[1], Frank A. Giordano[1,2,3], Jens Fleckenstein[1]

*[1] Department of Radiation Oncology, University Medical Centre Mannheim, Heidelberg University, Germany*

*[2] DKFZ-Hector Cancer Institute, University Medical Centre Mannheim, Mannheim, Germany*

*[3] Mannheim Institute for Intelligent Systems in Medicine (MIISM), Medical Faculty Mannheim, Heidelberg University, Mannheim, Germany*

**corresponding author*

Cornelius J. Bauer[1], *Ph.D.*

*Department of Radiation Oncology*

*University Medical Centre Mannheim, Heidelberg University*

*Email:* Cornelius.Bauer@umm.de

**Abstract:**

Introduction: Real-time dosimetry of surface doses in electron beams has not been widely established yet. Plastic scintillation detectors (PSD) promise high spatial resolution and real-time dosimetry with minimum perturbation of the radiation field. This study characterizes a novel PSD in an electron beam to determine its suitability for in vivo dosimetry.

Methods: Dual-channel Cherenkov radiation correction and dosimetric characterization of the PSD were investigated using reference ionization chambers. Percentage depth-dose curves, lateral profiles, and output factors were compared with reference ionization chamber measurements. Surface doses were measured on solid water and on an anthropomorphic phantom and were compared to ionization chamber and radiochromic film measurements.

Results: The investigated PSD demonstrated clinically acceptable linearity, dose rate independence, isotropy and reproducibility (total variation <2%). Dosimetric deviation in $R_{50}$ and $R_{80}$ were below 1.0 mm and lateral profiles agreed with a mean absolute error below 1.5%. Small field measurements were within 2% of the reference ionization chamber results. Surface dose measurements had mean relative deviations of 1.3% from ionization chamber measurements and 2.1% from radiochromic film measurements.

Conclusion: The PSD investigated in this study is suitable for clinically acceptable electron beam dosimetry and provides accurate dosimetric results for surface dose measurements. It has the potential to be used for real-time in vivo dosimetry.

## 1 Introduction:

Accurate dosimetry is essential in radiotherapy to ensure adequate therapeutic results while sparing healthy tissue[1, 2]. Dosimetry also plays a critical role in machine and patient-specific quality assurance as well as in vivo validation during irradiation. For electron beam radiotherapy, this becomes particularly important, as manual patient setup and beam collimation can markedly affect the total dose and the shape of the depth-dose curve[3, 4]. Hence, in vivo dosimetry can provide valuable information to verify a correct treatment delivery.

Approaches for in vivo dosimetry on the patient surface include the use of radiochromic films[5, 6, 7], MOSFET[8, 9] and diodes[10, 11]. While all these approaches have their strengths and weaknesses, a real-time, non-perturbative measurement that is independent of dose rate, energy and direction is not feasible with any of them[9, 10]. For absolute dosimetry, ionization chambers provide accurate results. However, they cannot be used for in vivo applications, as they perturb the beam during measurement and cannot be used on patient surfaces[12]. Hence, different detectors need to be investigated for electron in vivo dosimetry.

Plastic scintillation detectors (PSDs) have been introduced as active detectors that can be constructed in a compact design[13-15]. Commercially available PSDs are often used for high spatial resolution in stereotactic body radiotherapy (SBRT) applications[16] or in ultra-high dose rate (UHDR) conditions such as FLASH radiotherapy[17-20] due to their high temporal resolution. PSDs are claimed to be water-equivalent and thus to cause minimal disturbance to the beam during measurement. In addition, they provide real-time dose readout, which is crucial for in vivo dosimetry monitoring. The downside of PSDs is the need to correct for Cherenkov radiation, which leads to dosimetric uncertainties and a limited signal-to-noise ratio[21, 22].

The Blue Physics Model 11 plastic scintillation detector (SD, Blue Physics, Tampa, USA) is a novel PSD that offers high spatial and temporal resolution[23]. Characterization in a photon beam showed promising dosimetric results[24]. However, no characterization in electron beams has been published yet. Furthermore, the use of minimally perturbative PSDs for surface dose measurements has not yet been investigated.

In the present study, we characterized SD for the first time in an electron beam and evaluated its potential for accurate electron beam dosimetry. For usage in vivo, we evaluated the impact of the SD on the resulting dose distribution. Finally, we investigated the accuracy of surface dose measurements with SD and doses underneath a clinically realistic bolus material.

## 2 Materials and Methods

### 2.1 Detector and Experimental Setup

SD consists of a plastic scintillator, an optical fiber and an acquisition unit. The included software (BlueSoft) was used for data acquisition. Fig. 1 illustrates the detector composition. SD uses a dual-channel approach to correct for Cherenkov light that inevitably arises from partial irradiation of the optical fibers. A second, nearly identical fiber exclusively detects this Cherenkov light, which is then subtracted from the sensor signal using the so-called adjacent-channel ratio (ACR). This quantity corresponds to the Cherenkov light ratio used for other PSDs[25].

Manufacturer specifications indicate a sensitive volume of 0.785 mm$^3$ (a cylinder 1 mm long and 1 mm in diameter) and a minimum integration time of 300 µs. All measurements in this study were performed with an integration time of 750 µs. For evaluation, averaging was used to achieve an effective temporal resolution of 3 ms.
All measurements were performed on a medical linear accelerator (Linac, Synergy, Elekta AB, Stockholm, Sweden) with four electron beam energies (6, 8, 10, 12) MeV, at a dose rate of 400 MU/min unless stated otherwise. Electron applicators of 6 × 6 cm² or 10 × 10 cm² were used, depending on the required field size. Further beam collimation was performed via rectangular Rose's Metal (MCP-96) inserts. Source-surface distance (SSD) was 100 cm, and the measurement depth was chosen to be the reference depth $z_{ref}$, at the surface or below 1 cm of bolus material, depending on the experiment[26].

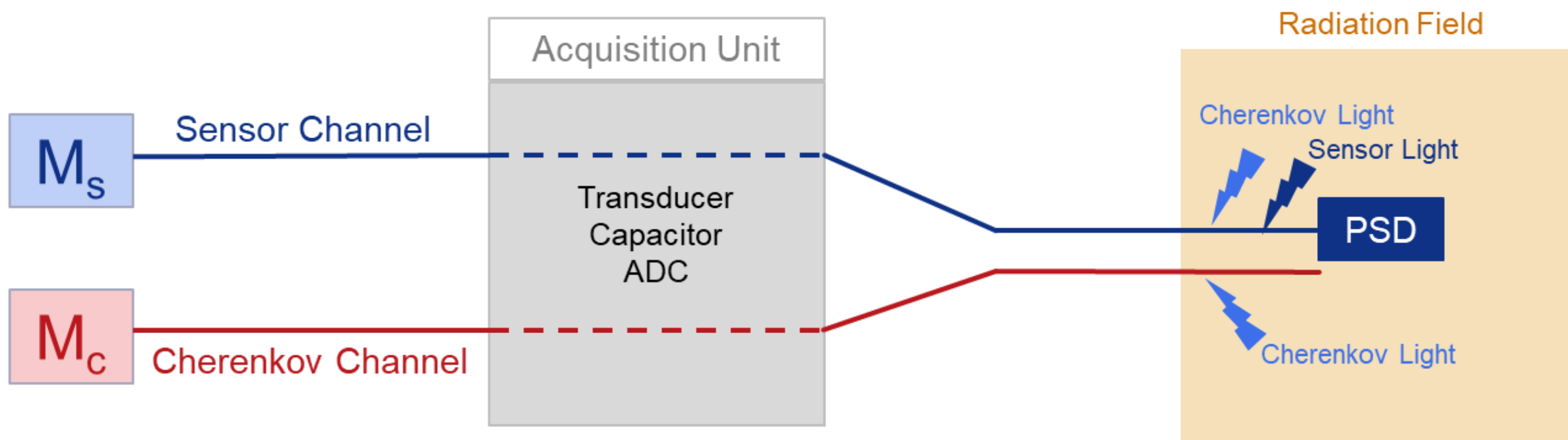

*Fig 1: Schematic detector setup of SD. Light generated in the plastic scintillator, as well as Cherenkov light, is acquired. Using a secondary channel that only measures Cherenkov light enables sensor light isolation.*

Characterization measurements were performed using a water phantom (MP3-XS, PTW, Freiburg, Germany) and a water-equivalent RW3 slab phantom (PTW, Freiburg, Germany).

Radiochromic films (RF, Gafchromic EBT-XD, Ashland Specialty Ingredients, NJ, USA) were evaluated 24 hours after irradiation using an in-house standardized protocol. 16-bit red-channel data, $R$ was used for color-to-dose calibration using data from an Advanced Markus ionization chamber (AM, Type 34045, PTW, Freiburg, Germany) and a regression with the function

$$D_{RF}\,(R) = a + \frac{b}{\log R - c}. \tag{1}$$

Model parameters were $a = -27.01\ \mathrm{Gy}, b = 55.44\ \mathrm{Gy}, c = 8.75$. The mean absolute relative error was 3.0%, $R^2 = 0.9994$ [27].

Whenever the arithmetic mean µ and standard deviation σ of a measurement are provided, the sample standard deviation was determined out of n= 3 repeated dose measurements $M_i$ with an identical measurement setup

$$\mu(M) = \frac{\sum_{i=1}^{n} M_i}{n}, \tag{2}$$

$$\sigma(M) = \sqrt{\frac{1}{n-1} \sum_{i=1}^{n} (M_i - \mu(M))^2}. \tag{3}$$

**2.2.1 ACR Measurement**

To determine the detector-specific ACR value, the method recommended by the vendor was used. For field sizes from 2 × 2 cm² to 20 × 20 cm², the dose equivalent to 200 MU in a 10 × 10 cm² field was determined using AM. The theoretical signal in the sensor channel $M_S$ can be expressed in terms of the Dose $D$, the calibration factor $f_{cal}$, the ACR, and the signal in the Cherenkov channel $M_C$:

$$M_S = \frac{D}{f_{cal}} + ACR \times M_C. \quad (4)$$

Measuring $M_S$ and $M_C$ for each field size three times (n= 3), the ACR was determined as the slope of a linear regression. The ACR-corrected signal used for all further analysis was denoted as

$$M = M_S - ACR \times M_C. \quad (5)$$

### 2.2.2 Dosimetric Characterization

First, dose linearity was determined over the range of 1 to 1000 MU in 15 logarithmically spaced increments. For this purpose, the following MUs were delivered and the detector reading $M$ recorded: The coefficient of variation

$$CV_{lin}\left(\frac{M}{MU}\right) = \frac{\sigma(M/MU)}{\mu(M/MU)} \quad (6)$$

was determined.
With a linear relation from energy dose $D$ as measured by AM and $M$, calibration was performed by combining equations (4) and (5):

$$D = f_{cal} \times M \quad (7)$$

For this calibration, repeated measurements with both SD (n= 3) and AM (n= 3) were acquired in a 10 × 10 cm$^2$ field at SSD= 100 cm for all electron beam energies. $f_{cal}$ was the mean of all energies. Dose rate $\dot{D}$ dependence was measured from 50 to 400 MU/min and calculated as the standard deviation of the measured dose $\sigma_{\dot{D}}(D)$ at a given dose rate divided by the mean dose at the maximum dose rate. The short-term repeatability was determined by repeatedly (n= 10) delivering a 10 × 10 cm$^2$, 10 MeV field and its standard deviation $\sigma_{st}$ was determined.
Finally, the isotropy of SD was determined by repeatedly (n=3) irradiating the detector from k=4 directions using all available beam energies. For this purpose, the detector was rotated in 90° increments using external markers. The CV(k) of the isotropy was determined.

### 2.3 Percentage Depth Dose Curves and Transversal Profiles

Percentage depth-dose curves (PDDs) were acquired using SD, AM and a Semiflex3D ionization chamber (SF, Type 31021, PTW, Freiburg, Germany). Ionization-to-dose conversion was performed according to TG-51 protocol[26]. PDDs in a 6 × 6 cm$^2$ field were measured for each energy from depth z= 80 mm to the water surface z= 0 mm.
Analysis of the PDDs was based on the depth of the maximum dose ($R_{max}$) and the depths of the 80/50% dose fall-off ($R_{80}$ and $R_{50}$). To investigate the PDD close to the surface, the PDD at 1 mm and 10 mm was evaluated.
To evaluate the potential for in vivo dosimetry, the perturbation of SD in the beam was investigated by acquiring a PDD with AM natively and with SD positioned at the water surface.

Transversal profiles were measured using SD and SF without ionization-to-dose conversion in depths of 0 cm and 1 cm. SF was positioned at the effective point of measurement using the TRUFIX detector positioning system (PTW, Freiburg, Germany). SD was positioned to have the center of the detector at the water surface for z= 0 mm. For all energies, a 6 × 6 cm$^2$ field was measured inline (IL) and crossline (CL) with a scan length of 12 cm. For evaluation, the mean absolute error (MAE) and mean error (ME) were considered across the scan length.

### 2.4 Small Field and Output-Factor Measurement

Correction free measurement of small fields was investigated by comparing the output factor (OF) curve. The central dose at 1 cm depth for field sizes from 3 × 3 $cm^2$ to 7 × 7 $cm^2$ was measured (n=3) using SD and AM. Normalization to the mean dose of the 7 × 7 $cm^2$ field yielded the OF. Gaussian error propagation was used to estimate the standard deviation of the OF for SD.

### 2.5 Surface Dose Measurement

The following surface dose measurements were performed: First, SD measurements in a 6 × 6 $cm^2$ radiation field were compared to RF and AM at the surface of the RW3 slab phantom and below 1 cm of bolus material (Bolx-I, Action Products, Hagerstown, MD, USA). Second, the same measurement was performed using a rectangular field collimation (3 × 10 $cm^2$). Additional collimator rotations of 45° and 90° were used to estimate the dosimetric error caused by the Cherenkov correction (Eq. 5). In this context, 0° collimator rotation represents the minimum and a collimator rotation of 90° represents the maximum amount of optical fiber in the beam. Finally, the dose at the surface and below 1 cm bolus material was measured with SD attached to an anthropomorphic thorax phantom (002LFC, Sun Nuclear, Norfolk, VA, USA). Due to the curved surface, RF was used as a dosimetric reference for this setup. All surface dose measurements were performed using 600 MU for all available energies. The experimental setup for surface dose measurements is shown in Fig. 2.

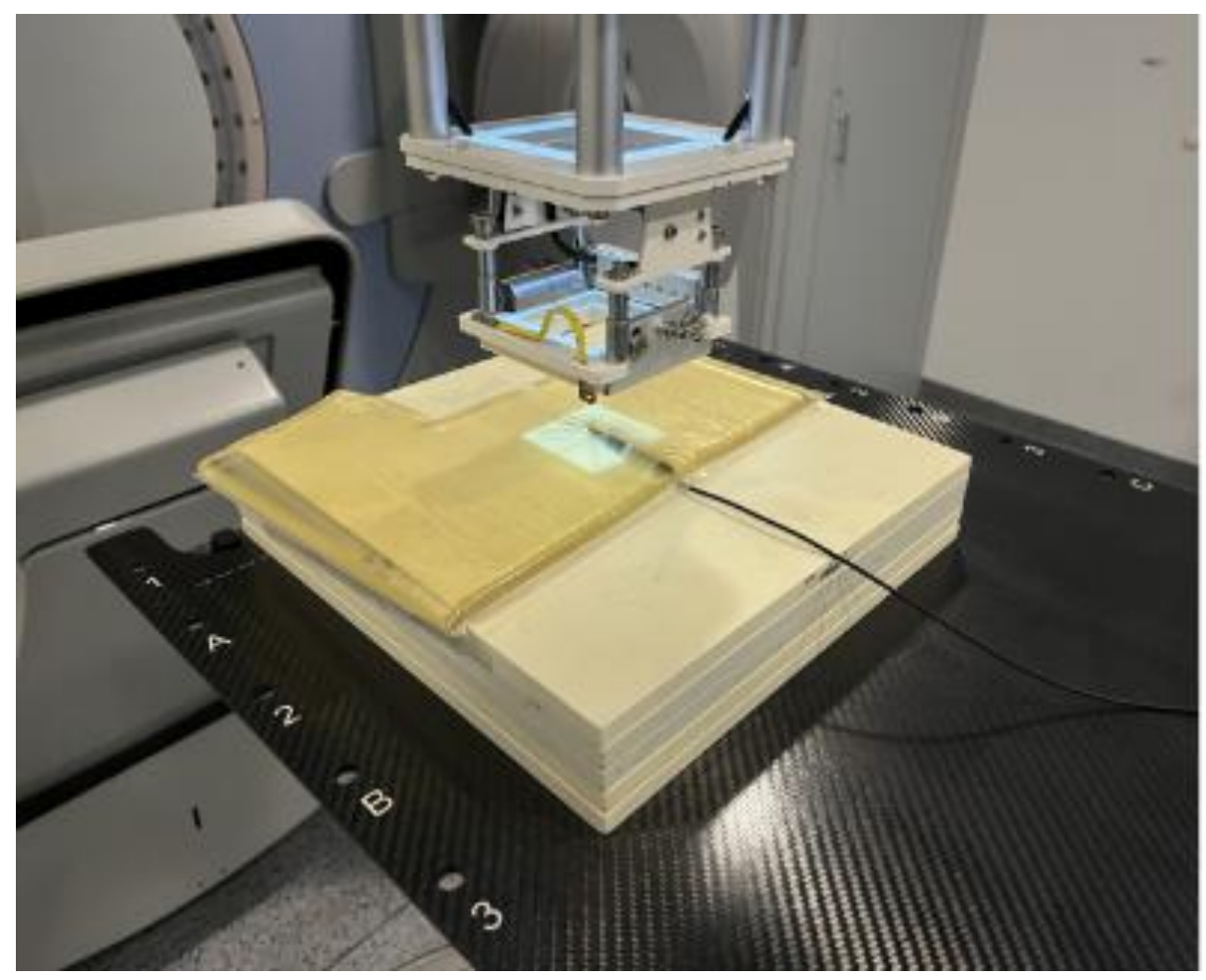

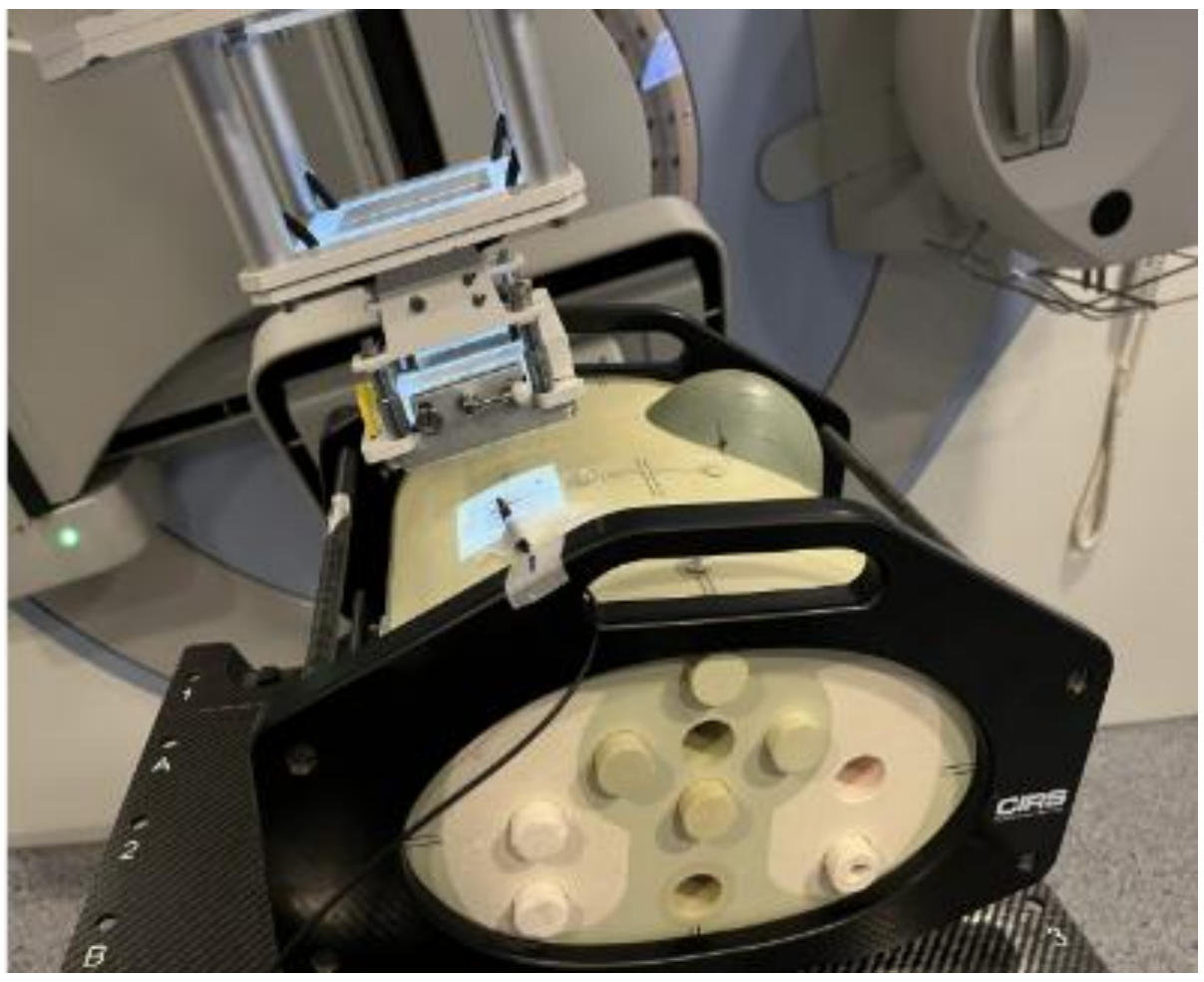

*Fig. 2: Experimental setup for surface dose measurements. Left: RW3 slab phantom with 1 cm bolus material. Gantry angle 0°. Right: Anthropomorphic phantom without bolus material, gantry angle 345°. SD is placed on the surface or right underneath bolus.*

## 3. Results

### 3.1 ACR and dosimetric characterization

The ACR was (0.916 ± 0.013), which is in line with the specifications by the manufacturer. This value was used to isolate the dose-proportional SD signal for all subsequent measurements.

The results obtained during detector characterization are shown in Tab. 1. Concerning deviations, the standard deviation is shown. All relevant properties were within 1.5%. A squared summation of all individual standard deviations (assuming independence) yielded a total standard deviation of 1.8%.

*Tab. 1: Detector characterization. For each detector property, the measurement procedure and the obtained value or standard deviation are given.*

| **Property** | **Measurement** | **Value/Standard deviation** |
| --- | --- | --- |
| ACR | ACR measurement using different field sizes | 0.916 ± 0.013 |
| $f_{cal}$ | Calibration against a reference ionization chamber (AM) | (2.29 ± 0.01) cGy/µC |
| Linearity $CV_{lin}$ | Detector signal from 1 to 1000 MU | 1.3% |
| Dose rate dependence | Detector reading for 100 MU at varying dose rates (50-400 MU/min) | 1.0% |
| Repeatability $\sigma_{st}$ | Detector reading for the same delivery settings (10 × ) | 0.3% |
| Isotropy CV(k) | Detector signal given rotation around the central axis | 0.8% |

### 3.2 Depth-Dose Curves and Water Equivalence

A comparison of PDD measured by SD and AM is shown in Fig. 3. For all energies, the PDDs align well. There is no AM data for depths < 1.3 mm due to the protection cap in water and the effective point of measurement[28, 29]. Tab. 2 summarizes the range and dosimetric differences between all detectors. $R_{80}$ and $R_{50}$ differ by less than 1.0 mm for each detector. $R_{max}$ shows a maximum deviation of 2.2 mm between SD and SF. At the depths of 1 mm and 10 mm, the maximum deviation is 1.0% between SD and SF, and 1.4% between SD and AM. The maximum deviation between AM and SF is 1.5%.

For in vivo dosimetry, SD within the radiation beam must not relevantly change the dose distribution within the patient. Comparing PDD measured with AM natively and with SD passively placed in the beam, a shift towards smaller ranges was observed. An analysis of $R_{50}$ and $R_{80}$ indicated shifts of 0.6 ± 0.4 mm, in line with the SD thickness.

*Tab. 2. Comparison of PDD measurements based on range parameters for SD, AM, and SF. Comparison of the PDD at 1 mm and 10 mm.*

**Incident electron beam energy**

*Fig. 3: PDD comparison measuring with SD (solid line), AM (dashed line) and SF (dotted line). Right: Overview from 0 to 80 mm. All PDDs align within 1mm. Left: Detailed view of first 30 mm. SF and SD align within 1% up to the surface. AM shows slight deviations close to the surface for 12 MeV.*

| Range | 6 MeV | 8 MeV | 10 MeV | 12 MeV |
|---|---|---|---|---|
| **$R_{max}$ (mm) (SD /AM/SF)** | 12.8/12.4/12.5 | 18.2/17.4/16.0 | 20.2/20.9/21.5 | 25.0/23.9/23.0 |
| **$R_{80}$ (mm) (SD /AM/ SF)** | 19.6/19.9/20.0 | 26.3/26.4/26.5 | 32.1/32.9/33.0 | 39.4/40.4/40.0 |
| **$R_{50}$ (mm) (SD /AM/SF)** | 23.9/23.9/24.0 | 31.5/31.9/32.0 | 38.4/39.4/39.0 | 47.2/47.4/48.0 |
| **$PDD_{1mm}$ (%) (SD/AM)** | 80.2/80.4 | 82.6/83.3 | 85.0/85.4 | 87.5/88.9 |
| **$PDD_{10mm}$(%) (SD/AM)** | 98.0/97.9 | 94.5/95.6 | 93.2/93.8 | 94.5/95.5 |

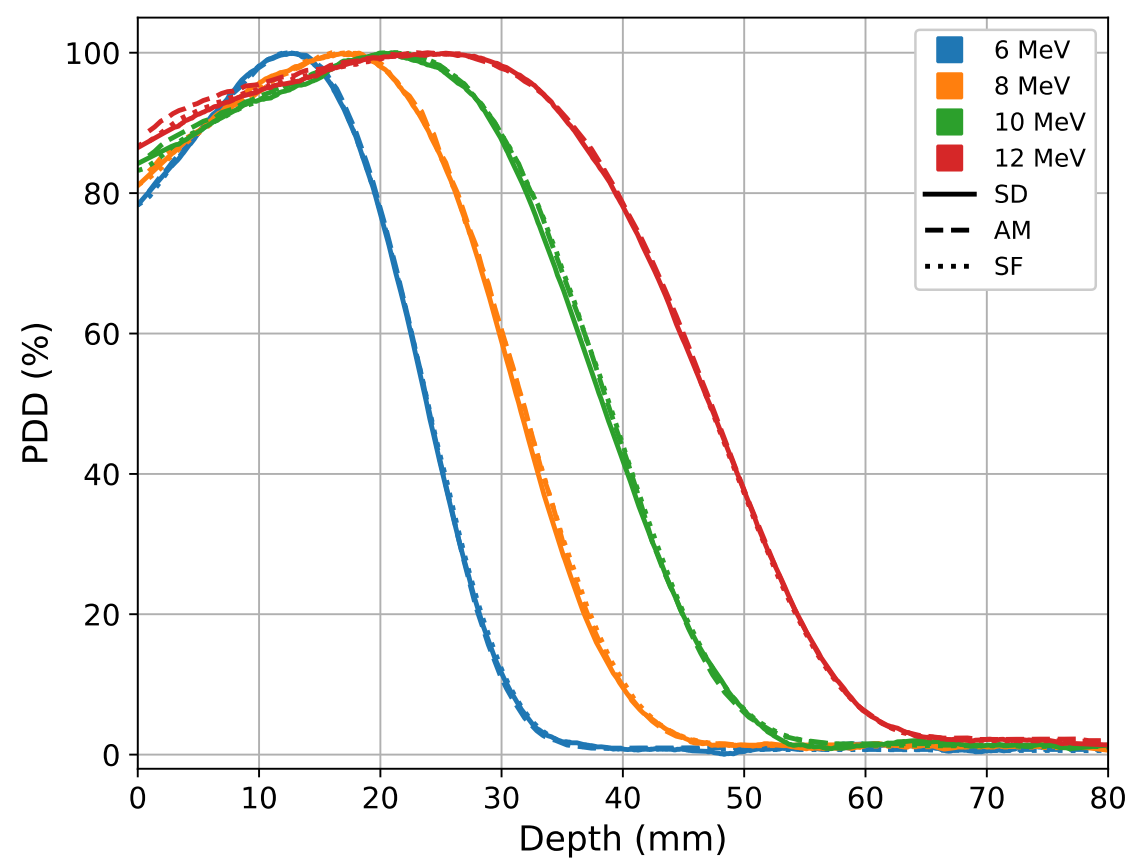

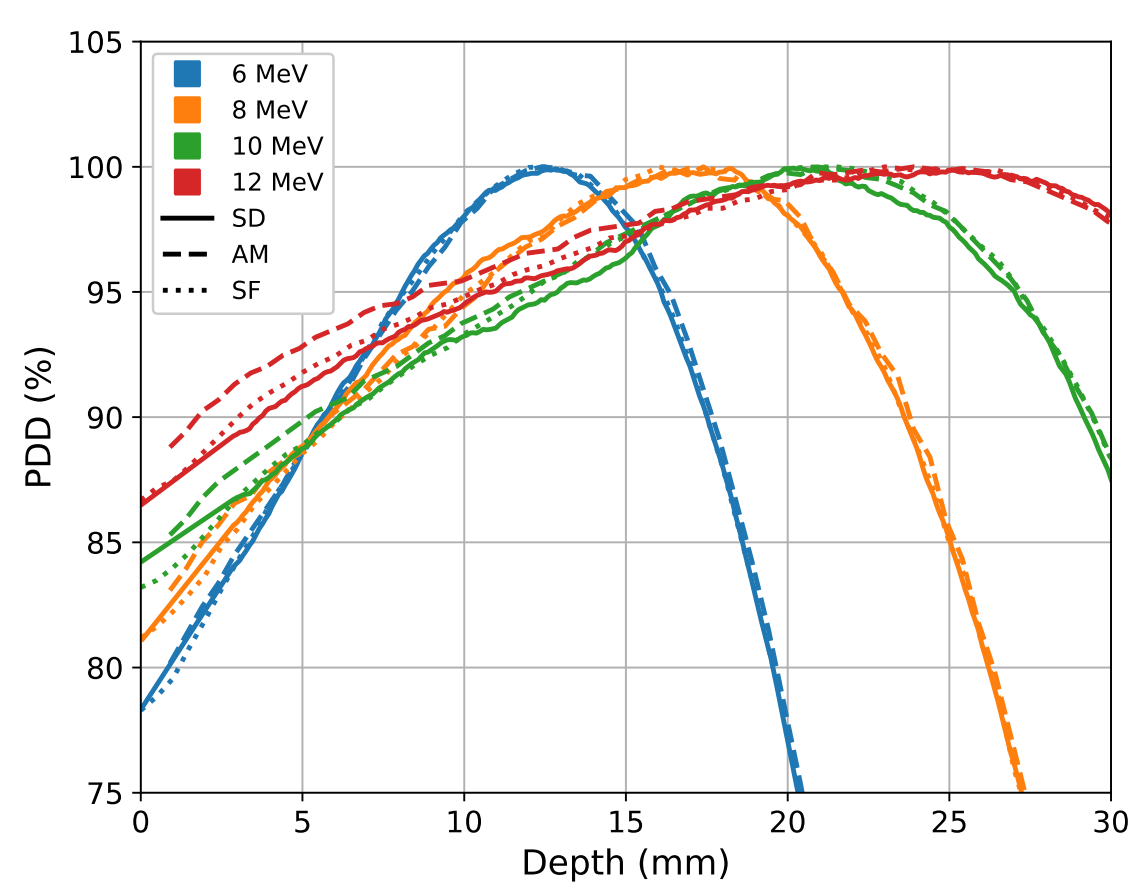

### 3.3 Transversal Beam Profiles

The differences in profile measurements between SD and SF are summarized in Tab. 3 using MAE and ME. Across the different energies, MAE ranged from 0.5% and 1.5%. MAE values for IL profiles were between 0.5% and 0.8% higher than in CL profiles. The ME ranged from -0.7% to 0.9% indicating no systematic shift between the measurements.

*Tab. 3: Mean absolute error (MAE) and mean error (ME) for profile measurements comparing SD and ionization chamber. Comparison at different energies for measurement depths of 0 cm and 1 cm.*

| | Incident electron beam energy | | | |
|---|---|---|---|---|
| **Direction** | **6 MeV** | **8 MeV** | **10 MeV** | **12 MeV** |
| **IL (0 cm)** | 1.4%/0.9% | 1.5%/0.8% | 0.9%/-0.4% | 0.7%/-0.2% |
| **CL (0 cm)** | 0.7%/-0.2% | 0.7%/-0.5% | 0.7%/-0.2% | 0.9%/-0.7% |
| **IL (1 cm)** | 1.1%/0.6% | 1.0%/0.5% | 0.9%/-0.3% | 0.9%/-0.4% |
| **CL (1 cm)** | 0.6%/0.0% | 0.5%/-0.3% | 0.6%/0.0% | 0.6%/0.0% |

Fig. 4 shows representative profiles (6 MeV and 12 MeV in CL direction). Visually, the profiles aligned both on the surface and in 1 cm depth. The difference plots underlined the noise-like nature of the differences.

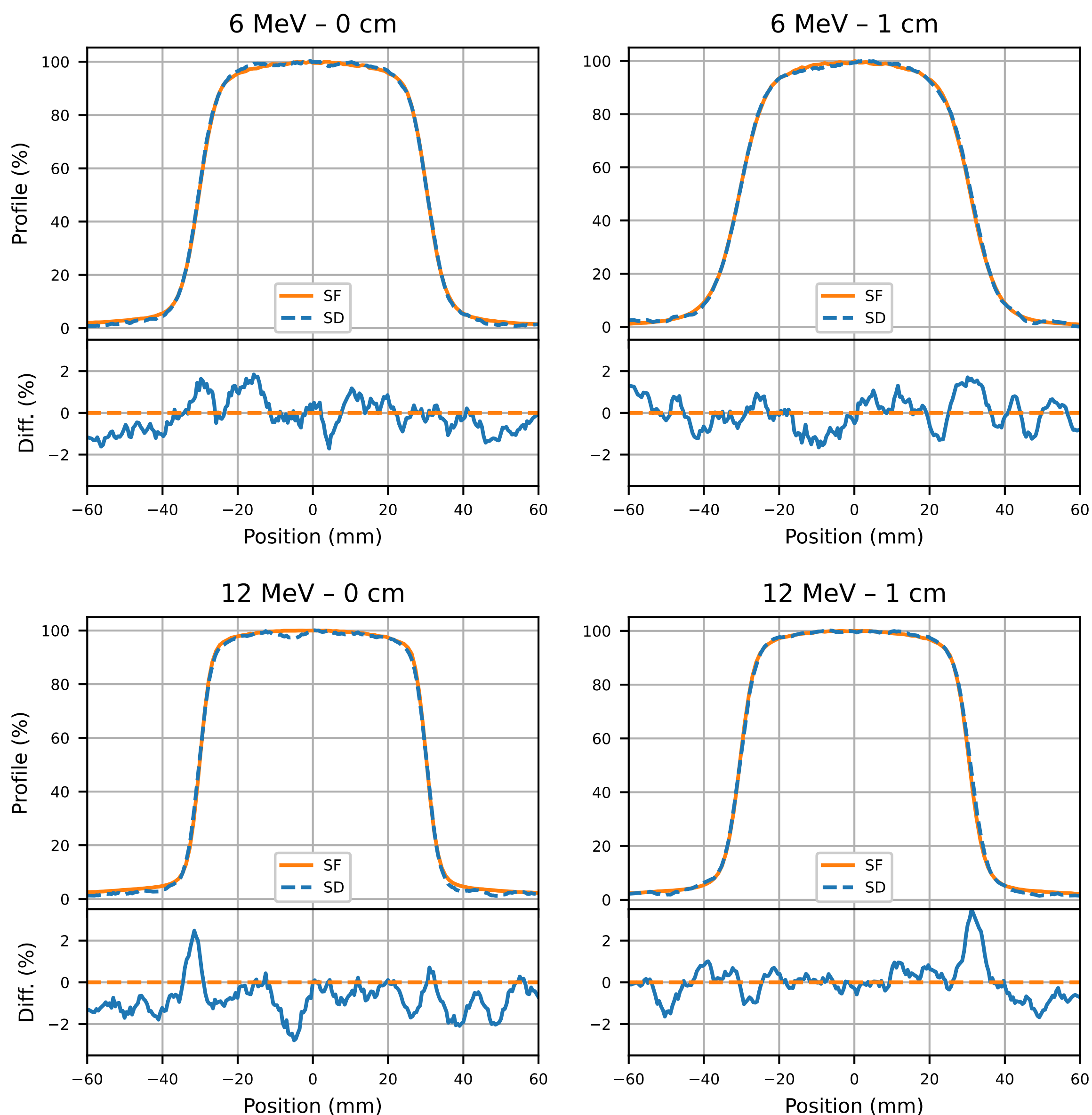

*Fig. 4: Representative profiles of 6 MeV and 12 MeV electrons in CL direction on the surface (0 cm, left) and in 1 cm depth (right). SD and SF data are shown with the difference plot below the profile. Visually, the profiles aligned and the differences were noise-like without trends or shifts. No difference above 3% was observed.*

### 3.4 Small Field and Output-Factor Measurements

For the use of SD in small fields, SD should measure the output factor curves without correction. Fig. 5 shows the OF curves for all energies measured with SD and AM. For SD, the mean ±1 standard deviation is displayed, while the standard deviation for AM is not displayed since it was <0.1% for all measurements. For field sizes from 3 × 3 cm² to 7 × 7 cm², SD and AM align very well with a mean

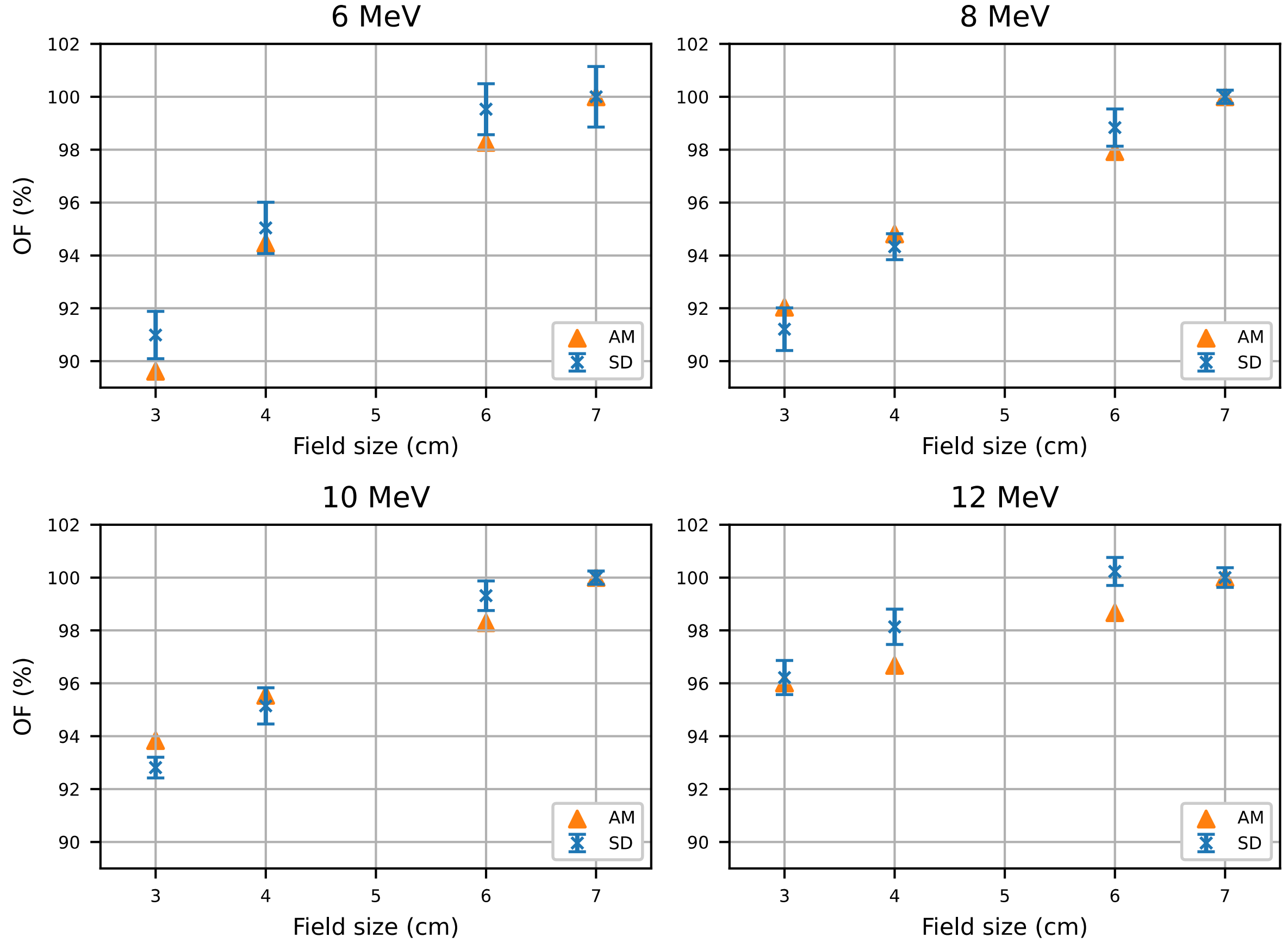

*Fig. 5: Output-factor curve for 6 MeV to 12 MeV measured with AM, and SD. Marker displays mean OF ± 1 standard deviation σ (n= 3). AM and SD align well and indicate comparable OF curves.*

deviation of (0.4 ± 0.8)% over all energies.

### 3.5 Surface dose measurements

Surface dose measurements are shown in Tab. 4. In the RW3 phantom and rectangular 6 × 6 cm$^2$ field, the MAE of the SD is below 1.5% when comparing to AM, SF, or RF. For the 3 × 10 cm$^2$ field, the MAE is 1.2% relative to AM, 1.3% relative to SF, and 2.1% relative to RF. As expected, the dose increased with energy and depth. Collimator rotations of the elongated rectangle yielded an MAE of 1.1% between scintillator measurements, indicating the effect of the Cherenkov correction. For measurements on the anthropomorphic phantom, the relative MAE was 1.4%. Over all RF measurements, the standard deviation of all percentage deviations was σ=2.7% (n=72).

*Tab. 4: Dose measurement deviation (in percent) of SD from different detectors on the surface and underneath a 1 cm bolus. All measurements were performed with a minimum amount of cable in the radiation field (collimator 0°). For RW3, SD was compared to AM, SF, and RF using a square field and a rectangular field. For the anthropomorphic phantom, RF served as the reference. For all RF measurements, the mean ± standard deviation is displayed.*

| | | Incident electron beam energy | | | |
|---|---|---|---|---|---|
| **Phantom/ field size** | **Reference** | **6 MeV (0 cm / 1 cm)** | **8 MeV (0 cm/ 1 cm)** | **10 MeV (0 cm / 1 cm)** | **12 MeV (0 cm / 1 cm)** |
| **RW3 6 × 6 cm²** | **AM** | -1.2/0.7 | -1.0/1.2 | -0.3/1.5 | -0.8/1.5 |
| | **SF** | -0.2/-2.8 | -0.8/-0.4 | -0.7/0.7 | -1.7/0.8 |
| | **RF** | 2.4±1.1/ -1.5±2.0 | 1.3±1.1/ 0.0±0.9 | -0.3±2.7/ 1.8±1.0 | 2.0±2.8/ 2.9±1.1 |
| **RW3 3 x 10 cm²** | **AM** | -2.9/-1.8 | -0.9/0.3 | 0.4/0.4 | 0.6/1.9 |
| | **SF** | 0.0/1.0 | 1.2/-1.8 | 2.2/-1.6 | 2.5/-0.3 |
| | **RF** | -3.5±2.3/ 1.6±2.7 | -1.1±2.5/ -1.6±1.6 | 0.8±1.7/ -4.2±0.6 | -1.3±1.0/ -3.8±2.7 |
| **anthr. Phantom 6 × 6 cm²** | **RF** | -0.5±1.6/ 0.3±1.4 | -3.2±2.6/ -0.7±1.7 | -0.1±2.3/ -2.2±1.2 | 3.5±1.6/ -0.2±1.0 |

## 4. Discussion

### 4.1 ACR and Dosimetric Characteristics

The determination of the ACR was crucial for correcting for Cherenkov radiation and providing a detector signal proportional to dose. This is the case for all PSDs or scintillation detectors in general[21]. The observed value of 0.916 was, as expected, close to 1 and comparable to the literature on PSDs[25, 30, 31]. In a 10 × 10 cm² square field, Cherenkov radiation contributed approximately 50% of the total signal. This means the Cherenkov contribution needed to be corrected adequately for a reliable dose measurement. Since the vendor recommendations[23] indicated a Cherenkov contribution of less than 50% of the total signal, this translated into fields shorter than 10 cm. Hence, no larger fields were considered in the present study. For in vivo applications, this would be mostly avoidable by positioning the optical fiber at the shortest length of the field. Consequently, the relative importance can be minimized and potential errors avoided. Surface dose measurements (see section 4.3) confirm the adequacy of the Cherenkov correction using ACR.

The dosimetric properties of SD were evaluated in four areas: linearity, dose-rate dependence, repeatability and isotropy. All of them were individually better than 1.5% and resulted in a total standard deviation of 1.8% when adding all effects under the assumption of independence.

The clinically acceptable dose-linearity of SD allowed for calibration using a constant calibration factor. In the regression, larger MU-values were essential for the fit routine, and deviations for small MU-values were expected and observed in the experiments. The greater overall variation in the detector readings for small doses (<5 cGy) raises the question of whether these differences are due to the Linac or the detector. Preliminary tests performed on the same Linac indicated that an ionization chamber also showed a larger relative standard deviation for such small doses. This is in line with common recommendations regarding the commissioning and quality assurance[32]. For most practical applications, doses below 5 cGy are not considered relevant dose prescriptions for radiotherapy with static electron fields. Therefore, variations in the smaller MU-regime were not investigated further.

The performed experiments indicated a deviation of less than 1.0% in the detector reading across the range of clinically relevant dose rates. This does not necessarily mean that SD is fully dose-rate-independent. For electron dosimetry in conventional clinical settings, it was still considered acceptable. Das et al., using the same detector for photons, found a comparable dose-rate dependence[24]. Other PSDs were reported to have dose-rate dependences of 0.2 to 1.0%, indicating SD within this range[25, 31, 32]. Since most publications considered photon irradiation, a strict comparison is difficult. When used at ultra-high dose rates, the behavior would have to be investigated further.

Small rotations can occur, particularly during in vivo dosimetry, and hence the angular dependence of SD was considered. The angular dependence was below 1%, enabling the use in clinical scenarios.

Overall, the detector characterization was in line with the findings for SD with photons[24, 30, 31] and comparable to the performance of other PSDs[25, 33, 34].

**4.2 Depth-Dose Curve and Profile Measurement**

An investigation of PDDs and lateral profiles was used to determine two aspects: First, accurate measurements of a PDD and profile by SD further indicate independence of dose rate, as the effective dose rate varies over the PDD and profile. Second, proper correction for the Cherenkov effect can be evaluated, as the amount of Cherenkov radiation is expected to vary throughout the measurement.
In the present study, the PDD measured with SD matches the reference from AM and SF. Both visually and in $R_{80}/R_{50}$, the discrepancies were below 1 mm. The only larger discrepancy was observed in $R_{max}$. This could be attributed to the plateau region around the maximum dose and the generally noisy SD data. This made it difficult to determine the maximum dose and, hence, Rmax. Given that the other PDD characteristics, particularly the surface regions, aligned, this was acceptable. Regarding AM measurements, the alignment at the surface was slightly inferior. Since this measurement included a protective water cap and a different detector holder, it appears plausible that the slight deviations below 1 mm originate from these differences.
Any in vivo measurement should not disturb the dose delivery. In the surface dosimetry use case, an important question was the effect of the SD's placement on the patient's surface on dose delivery. This was tested by measuring PDD using AM once in standard conditions and once with SD hovering above the water surface. Between these two measurements, PDD was shifted by less than 1 mm, and no further disturbance was observed.
Since PDD showed no further impact, SD can be considered water-equivalent for dosimetric purposes. This means that, for a future in vivo application, only the additional depth needs to be considered, without explicit modelling of the detector.
High-fidelity measurement of PDDs was also reported in other publications on SD and other PSDs[24, 25]. The impact of the detector on the PDDs was, to the best of our knowledge, not considered in other publications; in particular, for SD, this was not previously shown. This was, however, essential to ensure that in vivo measurements on the patient surface do not influence the dose delivery.
Profiles were acquired at depths of 0 cm and 1 cm in a water phantom to mimic the measurement situation at the surface and under 1 cm of bolus material, which is a realistic application scenario[35, 36]. These profile measurements were evaluated based on mean (absolute) differences. Mean absolute errors below 1.5% were acceptable for the profile measurements, in particular as they were to be expected given the dosimetric characterization (see section 4.1). The mean error indicated that the difference between SD and ionization chamber profile measurements was noise-like and not systematic. In the 12 MeV CL measurement with one of the larger deviations, a dip of 2% was observed. This was not present in any other profile measurements. Here, it was hypothesized that a small decrease in the Linac output could not be corrected in the measurement setup. Within the dosimetric characteristics of SD, this could also be a statistical variation in the detector signal. This suggests that for a use in quality assurance, an output correction would be advisable, and a bi-directional measurement could be employed to mitigate such effects.
Notably, the measurements of profiles and PDD with SD relied on the detector's constant movement through the water phantom. While this was the case in the present study, it would be crucial to verify before use in quality assurance.
In combination, the good agreement of SD with the reference detectors across these varying relative dosimetric measurements serves as an indicator that, even at the air-medium interface, the different devices agree when consistently positioned at the effective measurement point. Previously published Monte Carlo simulations indicate comparable results and margins of error of approximately 2%[37,38].

#### 4.3 Small Field and Surface Dose Measurement

The determination of the OF curve in this study was used to ensure accurate dose measurement in small fields, as they occur in typical in vivo settings[39,40]. The decision to acquire the OF at a 1 cm depth beneath the clinical bolus material was motivated by the typical clinical treatment scenario. In an in vivo approach, SD could realistically be placed on top of the bolus or just underneath it on the patient's skin. Measurements in reference depth $z_{ref}$ are not possible in vivo.
The use in small fields is essential for advancing in vivo dosimetry, as treatment with square fields without further collimation is rare[39, 40]. In small fields, the alignment of the uncorrected SD measurement and AM was promising in this regard. While the consideration of the OF was less in-depth than in some more specialized publications[41], it was deemed sufficient for in vivo use to estimate the correct dosimetric delivery.
Surface dose measurements showed excellent agreement between SD and the references (AM, SF, and RF) for rectangular fields in the RW3 phantom and the anthropomorphic phantom. Here, the equivalence of the measurement setups was crucial, so, for the curved geometry of the anthropomorphic phantom, RF was used as a reference, even though it had a larger relative dosimetric error. To the best of our knowledge, the present study was the first to demonstrate that a PSD accurately measures surface doses, an essential aspect for in vivo dosimetry.

There was a single outlier between SD and RF of -5.6% in a single measurement for 10 MeV underneath bolus. As RF intrinsically showed a 2.3% standard deviation, this was considered a statistical variation that could occur, given the number of experiments[42]. A positioning error was ruled out by the symmetrical dose on the film. Since the agreement between the ionization chambers and SD was good, no further investigation was performed. The total variation between SD and film was below 5% when considering the mean of three measurements.

In a clinically relevant elongated rectangle, dosimetric accuracy was comparable to ionization chambers. By varying the SD's central axis orientation in the field, the Cherenkov correction was tested under realistic conditions. In the recommended setup, with a minimal optical fiber length in the treatment field, the Cherenkov contribution was around 30%, increasing to 55% in the worst-case scenario with the fiber fully inside the radiation field. This observation was consistent with the characterization measurements discussed above. Ideally, the Cherenkov correction via the ACR should yield only a dose signal, i.e., a collimator rotation should not change the dose signal. The observed 1% difference in the dose measurement underscored that the Cherenkov correction worked adequately and enabled SD to be used for absolute dosimetric measurements.

Measurements at the air-medium interface are not straightforward, even for electron radiotherapy. The question of stopping-power ratios and the effect of missing scatter needs careful evaluation. Previous literature on this topic indicated the principal feasibility of such measurements in both photons[43] and electrons[44]. In particular, RF and parallel-plate chambers, such as AM, appear to be adequate references when no interpolation or Attix chamber is available[4,42-45]. In this study, the data indicate adequate agreement between SD and these references. When considering the application of in vivo monitoring, the accuracy was sufficient to estimate the dose in the tissue.
Overall, SD demonstrated minor deviations from ideal dosimetry of less than 2%. Given that OFs, PDDs, and profiles are well represented, we anticipate that in vivo use will be feasible.

## 5 Conclusion

The BluePhysics plastic scintillation detector provides accurate dosimetry for electron radiation in conventional clinical settings. Depth-dose curves, lateral profiles, and output factors were in excellent agreement with standard reference detectors. Surface dose measurements in phantoms demonstrated excellent agreement with ionization chamber and film measurements. This enables accurate in vivo dose measurements across a wide range of clinical scenarios with therapeutic electron beams.